\documentclass[letter, traditabstract]{aa} 
\usepackage{graphicx}
\usepackage{txfonts}
\usepackage{natbib}
\authorrunning{Brunthaler et al.}

\begin{document}
   \title{Discovery of a bright radio transient in M82: a new radio supernova?}


   \author{A. Brunthaler\inst{1}
          \and 
          K.M. Menten\inst{1}
          \and 
          M.J. Reid\inst{2}
          \and 
          C. Henkel\inst{1}
          \and 
          G.C. Bower\inst{3}
          \and 
          H. Falcke\inst{4,5}
          }

   \institute{Max-Planck-Institut f\"ur Radioastronomie, Auf dem H\"ugel 69,
              53121 Bonn, Germany\\
              \email{brunthal@mpifr-bonn.mpg.de}
         \and
             Harvard-Smithsonian Center for Astrophysics, 60 Garden Street,
              Cambridge, MA 02138, USA
         \and
              UC Berkeley, 601 Campbell Hall, Astronomy Department \& Radio
              Astronomy Lab, Berkeley, CA 94720, USA  
         \and
               Department of Astrophysics, Radboud Universiteit
               Nijmegen, Postbus 9010, 6500 GL Nijmegen, the Netherlands
               \and
              ASTRON, Postbus 2, 7990 AA Dwingeloo, the Netherlands
             }

   \date{Received 2009 April 14; Accepted 2009 April 21}

 
  \abstract
  {In this Letter, we report the discovery of a new bright radio transient in
    M82. Using the Very Large Array, we observed the nuclear region of M82 at
     several epochs at 22 GHz and detected a new bright radio source in this
     galaxy's central region. We find a flux density for this flaring source
     that is $\sim 300$ times larger than the upper limits determined in 
     previous observations. The flare must have
     started between 2007 October 29 and 2008 March 24. Over the past year,
     the flux density of this new source has decreased from $\sim$100 mJy to
     $\sim$11 mJy. The lightcurve (based on only three data points) can be
     fitted better with an exponential decay than with a power law. Based on
     the current data we cannot identify the nature of this transient
     source. However, a new radio supernova seems to be the most natural
     explanation. With its flux density of more than 100 mJy, it is 
     at least 1.5 times brighter than SN1993J in M81 at the peak of its
     lightcurve at 22 GHz.}  


   \maketitle
%

\section{Introduction}

M82 is a nearby \citep[3.6 Mpc based on a Cepheid distance to M81
by][]{FreedmanHughesMadore1994} irregular (I0) galaxy with a very
active starburst in its nuclear region. It harbors many bright supernova
remnants in its central region, which have been studied extensively
for decades
\citep{MuxlowPedlarWilkinson1994,BeswickRileyMartiVidal2006,Fenech2008}. 
\citet{vanBurenGreenhouse1994} estimate a supernova rate of 0.1
year$^{-1}$. However, a new radio supernova has not been discovered. 
\citet{Singer2004} reported a supernova in M82 (SN2004am) that was
classified as type-II \citep{Mattila2004}, but it has not been detected at
radio wavelengths \citep{Beswick2004}.    

\citet{KronbergSramek1985} and \citet{KronbergSramekBirk2000} monitored 
the flux densities of 24 radio sources in M82 from 1980 until 1992. Most
sources (75\%) remained surprisingly constant. There is some controversy about
how the  fluxes of these compact radio sources can be stable.
Models of supernova remnants expanding into a dense medium may explain this
\citep{ChevalierFransson2001}. \citet{SeaquistStankovic2007} argue that the
radio emission could arise from wind-driven bubbles.  Studying the evolution
of a young source could be very important for understanding these models. 
The strongest source, 41.9+58.0, shows an exponential decay with a decay 
rate of $\tau_d$=11.9 years \citep{KronbergSramekBirk2000}. Another source, 
41.5+597, which was detected in 1981 with a flux density of $\sim$10~mJy, 
faded within a few months to a flux density below 1 mJy 
\citep{KronbergSramek1985,KronbergSramekBirk2000}. The nature of this 
strongly variable source was never clarified. 

Radio supernova are rare events. So far only about two dozen have been detected
\citep{Weiler2002} and most of them were quite distant and rather weak. This
makes it difficult to study them in great detail. One notable exception is 
SN1993J \citep{Schmidt1993} in M81, which has been studied extensively
\citep{Marcaide1997,Marceide2009, Bietenholz2001, Bietenholz2003, Perez2001,
 Perez2002, Bartel2002, Bartel2007}. Thus, the detection of a new nearby
supernova would be highly desirable. 

M82 is part of the M81 group of galaxies and it shows clear signs of tidal
interaction with M81 and NGC\,3077 \citep{YunHoLo1994}. This makes the M81
group an ideal system for studying galaxy interaction in great detail. With 
this
motivation, we have initiated a project to measure the proper motions of M81
and M82 with VLBI astrometry. We are observing M81*,  the nuclear radio source
in M81, bright water masers in M82, and three compact background
quasars. Based on our experience with measurements of proper motions in the
Local Group \citep{BrunthalerReidFalcke2005, BrunthalerReidFalcke2007}, we
expect a detection of the tangential motions of M81 and M82 relative to the
Milky Way within a few years. So far, we have observed at three epochs at 22
GHz with the High Sensitivity Array (including the Very Long Baseline Array,
the Very Large Array, and the Greenbank and Effelsberg telescopes). 
Here, we report the detection of a new transient source in M82 based on the
data from the NRAO\footnote{The National Radio
  Astronomy Observatory is a facility of the National Science Foundation
  operated under cooperative agreement by Associated Universities, Inc.} Very
Large Array (VLA).  


\section{Observations and data reduction}

M82 was observed with the Very Large Array (VLA) as part
of the High Sensitivity Array observation under projects BB229
and BB255 on 2007 January 28, 2008 May 03, and 2009 April
08. The total observing time in each epoch was 12 hours. We
used M81* as phase calibrator and switched between M81*, M82, and 3
extragalactic background quasars every 50 seconds
in the cycle M81* -- 0945+6924 -- M81* -- 0948+6848 -- M81*
-- M82 -- 1004+6936 -- M81*, yielding an integration time of $\sim$100
minutes on M82. We observed with two frequency
bands of 50 MHz, each in dual circular polarization.
The data reduction was performed with the Astronomical
Image Processing System (AIPS) and involved standard steps.
On 2007 January 28 and 2009 April 08, we used 3\,C48 as
flux density calibrator. M81* was used as a gain and phase calibrator and
one round of phase self-calibration was performed on M82.
Unfortunately, no flux density calibrator was observed on 2008
May 03. Here, we assumed a flux density of 150 mJy for the highly variable
source M81*. This value was chosen since it was in the range of typical values
at cm wavelenghts \citep[e.g.][]{BrunthalerBowerFalcke2006} and it yielded flux
densities for 0945+6924, 1004+6936, and 44.0+59.6 in M82 that were consistent
with their flux densities at the other two epochs.   

M82 was also observed on 2008 March 24 with the VLA at 22 GHz for 10 minutes
($\sim$6 minutes integration time on M82) in spectral line mode. 3C\,48 was
used as flux density calibrator, and 1048+717 was used as gain and phase
calibrator. A total bandwidth of 9.18 MHz was observed. We also analyzed the
archival VLA data of 2007 October 29 at 4.8 GHz, which is the latest available
data before our observations. Here, 3C\,286 was used as flux density
calibrator and 1048+717 was used as phase calibrator. Eighteen minutes of
integration time on M82 was spread over 1.5 hours.

\section{Results}

\begin{figure}[!h]
   \resizebox{0.5\textwidth}{!}
            {\includegraphics[bb=0.0cm 1.8cm 25.1cm 51.2cm,clip,
              angle=0]{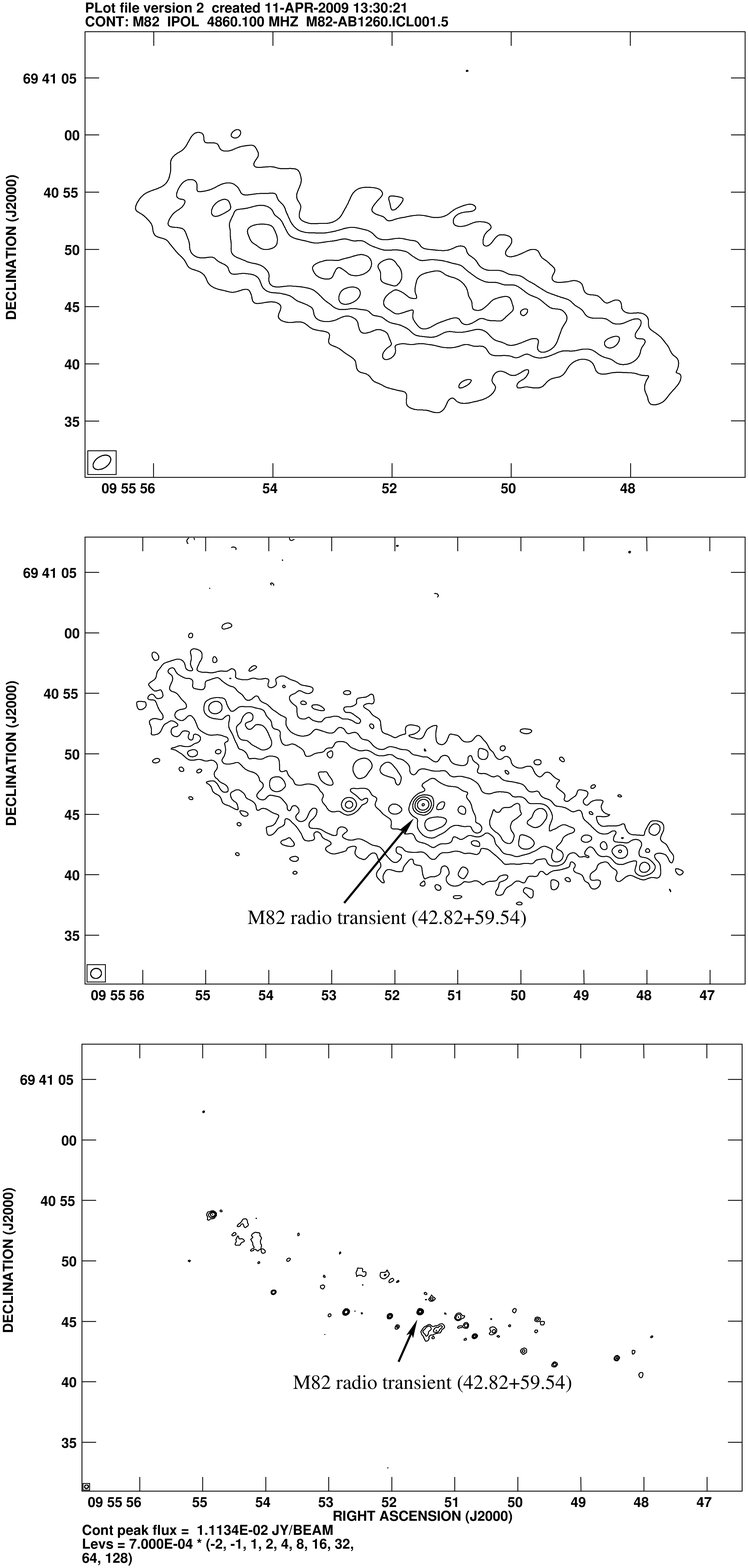}} 
   
      \caption{
        {\bf Top:} VLA B-configuration image at 4.8 GHz of M82 on 2007
        October 29. Contours start at 3 mJy and increase with factors of
        2. The beamsize is 1.7$\times$1.1 arcseconds with a position angle
        of -58$^\circ$. There is no detectable source at the position of our 
        new source. The peak flux in
        the image is 45 mJy~beam$^{-1}$.   
        {\bf Middle:} VLA C-configuration image at 22 GHz of M82 on 2008 May
        03. Contours start at 0.7 mJy and increase with factors of 2. The
        beamsize is  0.90$\times$0.83 arcseconds at a position angle of
        -73$^\circ$. The new source at position
        $\alpha_{J2000}$=09$^{h}$55$^{m}$51.551$^{s}\pm0.008^{s}$
        $\delta_{J2000}$=69$^\circ$40$'$45.792$''\pm0.005''$ is clearly
        visible, with a peak flux of 90 mJy~beam$^{-1}$. 
        {\bf Bottom:} VLA B-configuration image at 22 GHz of M82 on 2009 April
        08. Contours start at 0.7 mJy and increase with factors of 2. The
        beamsize is  0.32$\times$0.29 arcseconds with a position angle of
        -60$^\circ$. The new source is still clearly
        visible, but its peak flux density has decreased to 11 mJy~beam$^{-1}$.  
   }
         \label{Fig:detect}
   \end{figure}

\begin{figure}[!h]
   \resizebox{0.5\textwidth}{!}
             {\includegraphics[angle=0]{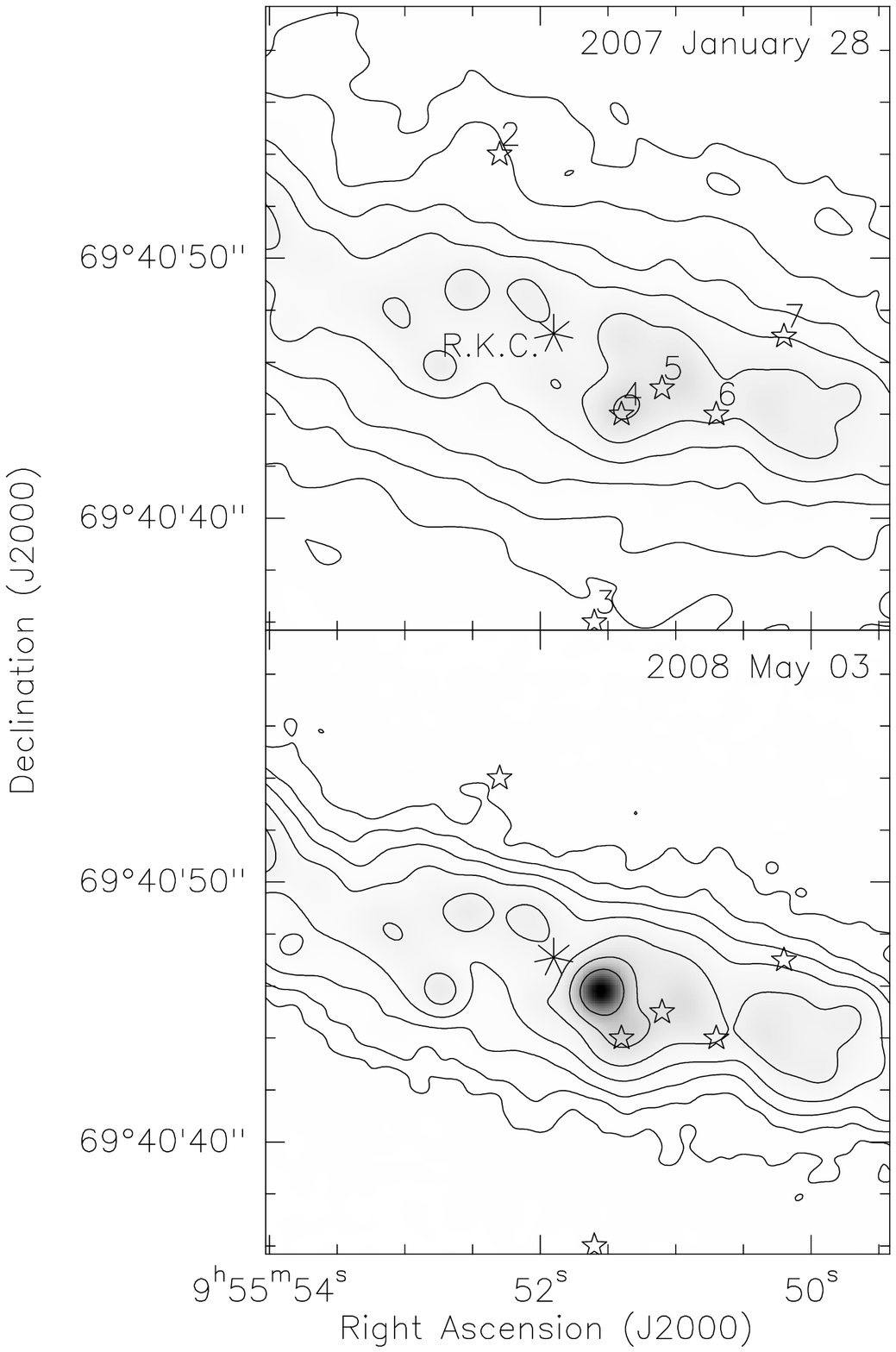}}
      \caption{Contours and grey scale represent 22.2 GHz images of the central
region of M82 taken on 2007 January 28 ({\bf top}) and 2008 May 03 ({\bf
bottom}). The new bright source is very conspicuous in the latter image. Contour
values are -4, 4, 8, 16, 32, 64, 128, 256, and 512 times 0.2 mJy~beam$^{-1}$,
roughly the (comparable) rms noise level in both images. The star symbols mark
the positions of the x-ray sources discussed by \citet{Matsumoto2001} and in
the top panel bear these authors' nomenclature. The asterisk gives the position
of the radio kinematic center (R.K.C.) of M82 as determined by
\citet{WeliachevFomalontGreisen1984}. Both images were restored with a
circular beam of 1.5 arcseconds. Since the observation on 2007 January 28 was
made in C configuration, there was much more data from short baselines, which
results in better sensitivity to extended emission.
   }
         \label{Fig:detail}
   \end{figure}

A new bright source was detected on 2008 May 03 (see Fig.~\ref{Fig:detect},
middle and Fig.~\ref{Fig:detail}, bottom). With it’s flux density of $\sim90$
mJy, it was clearly the brightest radio source in the field (at least five
times brighter than the second brightest source). Almost one year later, on
2009 April 08, the source had faded to a flux density of $\sim11$ mJy,
i.e. losing almost 90\% of it’s flux density (see Fig.~\ref{Fig:detect}, bottom). We
estimated its position to be
$\alpha_{J2000}$=09$^{h}$55$^{m}$51.551$^{s}\pm0.008^{s}$
$\delta_{J2000}$=69$^\circ$40$'$45.792$''\pm0.005''$. Following the
traditional convention of previous papers naming a source after the
position offset to $\alpha_{B1950}$=09$^{h}$55$^{m}$ and
$\delta_{B1950}$=69$^\circ40'$, our new detection would be 42.82+59.54. The
positions in the two observations are consistent and the uncertainty was
estimated by comparing the position of another clearly identified source
(44.0+59.6) with the positions from observations with the Multi-Element Radio
Linked Interferometer Network (MERLIN) by \citet{MuxlowPedlarWilkinson1994}.  

To constrain the start time of the flare, we reduced two earlier VLA 
observations of M82. The spectral line observation on 2008 March 24 had lower 
quality, since we observed with a narrow bandwidth and had only $\sim$6 minutes
integration time on M82. Nevertheless, we could easily confirm the detection
of the new source at the same position with a similar but slightly higher flux
density as in the 2008 May 03 observation. Next, we analyzed 4.8 GHz data from
2007 October 29. Here, no bright point source was discovered (see
Fig.~\ref{Fig:detect}, top),  so we conclude that the start of the flare was
between 2007 October 29 and 2008 March 24. However, it is possible that the
source could be highly self-absorbed early in its development and thus might 
have been detectable at high frequencies, but not at 4.8 GHz.  In this case, 
the flare might have started earlier, but not earlier than 2007 January 28  
(the day of our first 22 GHz observation, see Fig.~\ref{Fig:detail}, top).  
 
The comparision of flux densities from different epochs in M82 is difficult
due to the diffuse emission, in particular when comparing observations with
different spatial resolutions. In Table~\ref{tab:flux}, we list the flux
densities of the new source along with the flux densities of one additional
source in M82 (44.0+59.6), M81*, and three background quasars.

\section{Pre-flare observations at other wavelengths}
\citet{Matsumoto2001} present high-resolution (FWHM $\sim 0\rlap{.}''5$) x-ray
imaging of the central $1'\times1'$ (1.1 $\times$ 1.1 kpc) region of M 82
with the High-Resolution Camera (HRC) aboard the Chandra X-ray Observatory.
These images, taken on 1999 October 28 and 2000 January 20, show a total of 9
sources within this area, some of which are highly variable. They do
not find a counterpart to our new radio source. Neither do \citet{Kong2007} in
a total of 12 datasets of a similar sized region taken with the Chandra HRC and
Advanced CCD Imaging Spectrometer Array (ACIS-1 and -2) taken between 1999
September 20 and 2005 August 18. These authors also present Hubble Space
Telescope (HST) $H-$band ($1.6~\mu$m) imaging of the region with
the Near-Infrared Camera and Multi-Object Spectrometer (NICMOS) that, again,
does not show a counterpart to our variable radio source.

\citet{KoerdingColbertFalcke2005} observed M82 with the VLA at 8.4
GHz eight times between 2003 June and October in order to detect possible 
radio flares
from several ultraluminous X-ray sources (ULX). No significant emission was 
found at the position of our new detection above a noise level of 70 $\mu$Jy
at each epoch. A very sensitive eight-day integration of M82 at 5 GHz
wavelength was performed with MERLIN between 2002 April 1 and 28
\citep{Fenech2008}. The resulting 40 millarcsecond resolution images show no
significant emission at the position of our new source above the rms noise
level of $17~\mu$Jy~beam$^{-1}$. \citet{Tsai2009} present a 7 mm map from
the VLA from 2005 April 22, with no detection of more than 0.3 mJy at
the position of our source.  

\section{Discussion}
The most straightforward explanation for this new source is a new radio
supernova. A quantitative analysis of the lightcurve is difficult due
to the small number of data points (3) we obtained with different angular
resolutions, the complication of diffuse emission in  M82, and the uncertain 
absolute flux scale in our observation of 2008 May 03.  We extracted the flux d
ensities of our source by restricting the interferometer (u,v)-data to 
$>$30~k$\lambda$ to remove most of the extended emission 
(Table~\ref{tab:flux}).   

\begin{figure}
   \resizebox{0.5\textwidth}{!}
             {\includegraphics[angle=-90]{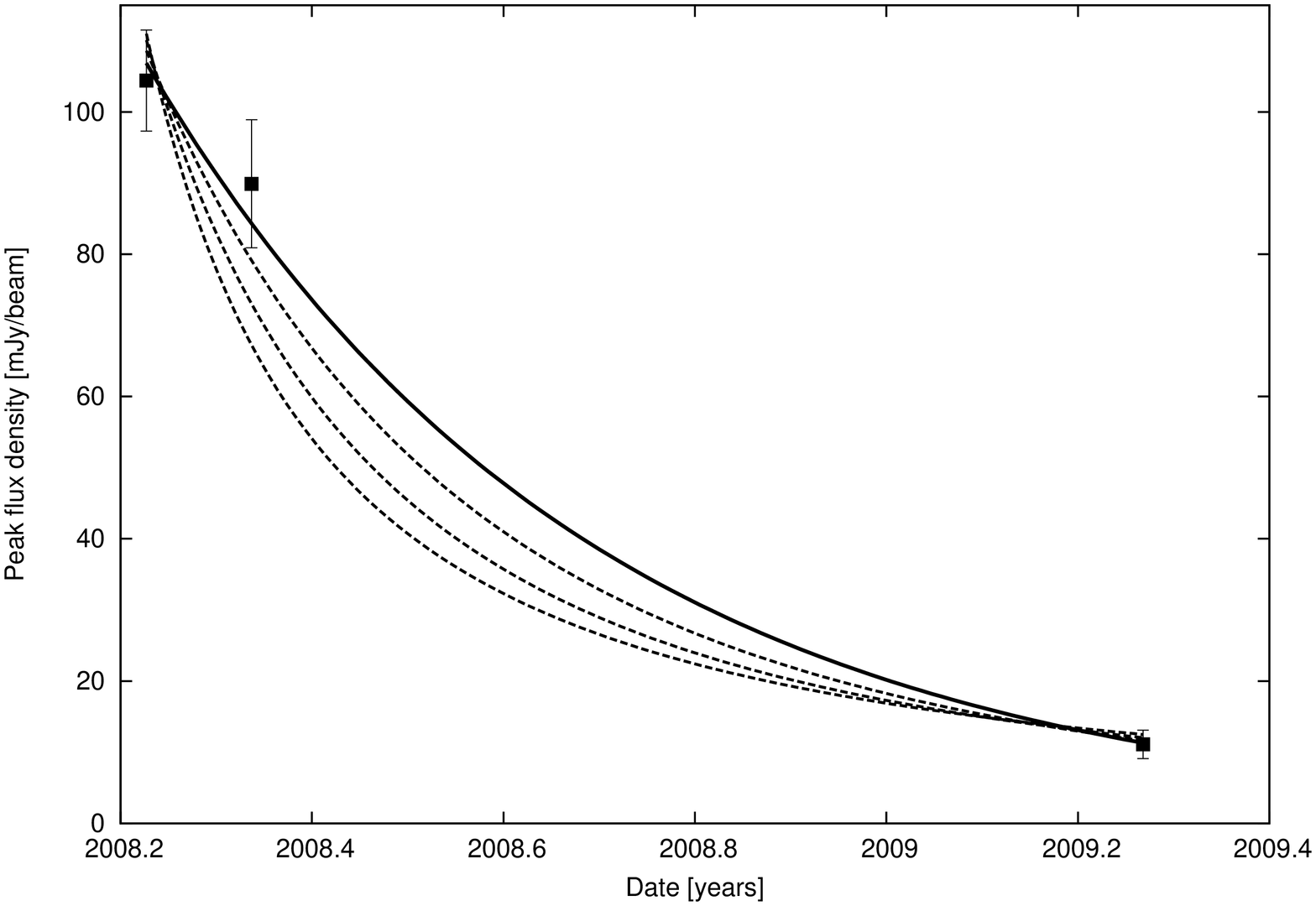}}
      \caption{Lightcurve of the new radio transient from our
        observations (squares). Also shown are an exponential-decay fit (solid
        line) and power-law fits (dashed lines), assuming different peak times 
        of the flare: 2008.0 (lower line), 2007.8 (middle line), and 2007.08 
        (upper line). 
   }
         \label{Fig:light}
   \end{figure}

We fitted the lightcurve of the source with an exponential decay:    
S(t)=S$_0$ e$^{(t_0-t)/\tau_{d}}$.
This yields a decay timescale of $\tau_d$=0.46$\pm$0.03 yr
(Fig.~\ref{Fig:light}). Since we do not know the exact time of the onset of the
flare, t$_0$, we get values of S$_0$ between 110 and 270 mJy for t$_0$ between
2008 March 24 and 2007 October 29. For the fit, we added in quadrature the
difference between peak and integrated flux density as one error estimate and 
systematic (flux density scale) errors of
5\% for the first and third epochs, and 10\% for the second epoch, where no
proper flux density calibrator was observed. The resulting $\chi^2_{pdf}$ was
0.5. Based on this fit, we predict that the source flux density will drop to 5
mJy by mid 2009 and 1.5 mJy in early 2010. We also fitted a simple exponential
decay to the 22 GHz lightcurve of SN199J in M81. Here we used the first year
of data after the peak in the lightcurve published in
\citet{WeilerWilliamsPanagia2007}. The fit yields $\tau_d$=0.72$\pm$0.09 yr,
indicating that the new source in M82 decays faster than SN1993J.  

A power-law fit (S(t)$\propto$(t$_0$-t)$^\alpha$) is not consistent with the
lightcurve of the new source in M82 ($\chi^2_{pdf}=5-8$, for t$_0$ between 
2007.8 and 2008). If one allows an earlier t$_0$ (e.g., if the source is highly
self-absorbed at lower frequencies), the power-law fit improves
($\chi^2_{pdf}=2$, for t$_0$=2007.08, one day after our 2007 January 28
observation).

The position of the new source is marginally consistent with the position of
the kinematic center of M82 \citep{WeliachevFomalontGreisen1984}. This raises
the possibility that, rather than a supernova, we could have detected a
flare from a supermassive black hole in the center of M82. Flares in AGN 
sources can often be fitted with exponential decays
\citep{ValtaojaLaehteenmaekiTeraesranta1999}.  For
example, a strong radio flare in 1999 in the Seyfert galaxy III\,Zw\,2 has a
decay rate of $\tau_d$=0.73 year \citep{BrunthalerFalckeBower2005}. 

Other explanations, such as luminous flares from quiescent supermassive black
holes induced by a close passage of a star that is torn apart by tidal forces, 
are also possible. However, such flares show a power-law decay with
$\alpha=-5/3$ \citep{EvansKochanek1989, Ayal2000, Gezari2009}, which is not
consistent with our lightcurve.

However, M82 has never shown evidence of a nuclear supermassive black hole
(which would be surprising for a small irregular galaxy).  Since the progenitor
for our flare showed no X-ray emission, a stellar or intermediate mass black 
hole is also not probable. Thus, based on the current data, a new radio 
supernova seems to be the most likely explanation. 

\begin{table}
\caption{Details of the detected compact sources in our VLA observations with  
statistical errors on the 22.2 GHz flux densities.}
\label{tab:flux}      
\centering           
\begin{tabular}{cccccc}   
\hline\hline               
Source & Date &  Peak Flux & Integrated Flux\\
     &        &  [mJy~beam$^{-1}$]     & [mJy]\\        
\hline                        
M82 transient & 24/03/2008 &  104.4$\pm$1.3 & 99.6$\pm$2.3 \\
(42.82+59.54)          & 03/05/2008 &  89.9$\pm$0.1  & 88.4$\pm$0.2 \\
          & 08/04/2009 &  11.1$\pm$0.1   & 9.2$\pm$0.2   \\

\\
44.0+59.6 & 28/01/2007 & 15.0$\pm$0.2  & 77.8$\pm$1.0  \\
          & 03/05/2008 & 12.5$\pm$0.1  & 11.0$\pm$0.2  \\
          & 08/04/2009 & 14.7$\pm$0.1  & 12.4$\pm$0.2  \\
\\
M81*      & 28/01/2007 & 71$\pm$1      & 72$\pm$1      \\
          & 03/05/2008 & 150           & 150           \\
          & 08/04/2009 & 195$\pm$3     & 179$\pm$5     \\
\\
0945+6924 & 28/01/2007 & 14.3$\pm$0.2 & 14.4$\pm$0.3  \\
          & 03/05/2008 & 14.8$\pm$0.1  & 14.7$\pm$0.2  \\
          & 08/04/2009 & 13.8$\pm$0.2  & 12.2$\pm$0.3  \\
\\
0948+6848 & 28/01/2007 & 22.5$\pm$0.2 & 24.3$\pm$0.3  \\
          & 03/05/2008 & 59.1$\pm$0.2  & 59.8$\pm$0.3  \\
          & 08/04/2009 & 63.6$\pm$0.4  & 56.3$\pm$0.6  \\
\\
1004+6936 & 28/01/2007 & 31.0$\pm$0.2 & 30.8$\pm$0.3  \\ 
          & 03/05/2008 & 30.9$\pm$0.2  & 30.4$\pm$0.2  \\
          & 08/04/2009 & 34.6$\pm$0.3  & 30.8$\pm$0.4  \\

\hline                                   
\end{tabular}
\end{table}

\begin{acknowledgements}
We thank the referee Dr. K. Weiler for critically reading the manuscript.
\end{acknowledgements}

\bibliographystyle{aa}
\bibliography{brunthal_refs}

%


\end{document}